\documentclass[%
 reprint,
 amsmath,amssymb,
 aps,
]{revtex4-2}
\usepackage{graphicx}
\usepackage{dcolumn}
\usepackage{bm}
\usepackage{xcolor}

\usepackage{multirow}

\newcommand{\ilm}{Univ Claude Bernard Lyon 1, CNRS, Institut Lumi\`ere Mati\`ere, F-69622, VILLEURBANNE, France}
\renewcommand{\selectlanguage}[1]{}
\DeclareUnicodeCharacter{2212}{-}
\usepackage{booktabs}
\begin{document}

\preprint{APS/123-QED}

\title{Thermoplasmonics of Gold-Core Silica-Shell Colloidal Nanoparticles under Pulse Illumination }

\author{Julien El Hajj}
\email{julien.el-hajj@univ-lyon1.fr}
\author{Gilles Ledoux}
\author{Samy Merabia}
\email{samy.merabia@univ-lyon1.fr}
\affiliation{\ilm}

\date{\today}
\begin{abstract}
Core-shell nanoparticles, particularly those having a gold core, have emerged as a highly promising class of materials due to their unique optical and thermal properties, which underpin a wide range of applications in photothermal therapy, imaging, and biosensing. In this study, we present a comprehensive study of the thermal dynamics of gold-core silica-shell nanoparticles immersed in water
under pulse illumination. The plasmonic response of the core-shell nanoparticle is described by incorporating Mie theory with electronic temperature corrections to the refractive indices of gold, based on a Drude–Lorentz formulation. The thermal response of the core-shell nanoparticles is modeled by coupling the two temperature model with molecular dynamics simulations, providing an atomistic description of nanoscale heat transfer. We investigate nanoparticles with both dense and porous silica shells (with 50\% porosity) under laser pulse durations of $100$\,fs, $10$\,ps, and $1$\,ns, and over a range of fluences between $0.05$ and $5$\,mJ/cm$^2$. 
We show that nanoparticles with a thin dense silica shell ($5$\,nm) exhibit significantly faster water heating compared to bare gold nanoparticles. This behavior is attributed to enhanced electron-phonon coupling at the gold silica interface and to the relatively high thermal conductance between silica and water. These findings provide new insights into optimizing nanoparticle design for efficient photothermal applications and establish a robust framework for understanding energy transfer mechanisms in heterogeneous metal dielectric nanostructures.

\end{abstract}

\maketitle

\section{Introduction}
Metallic nanoparticles, particularly gold nanoparticles, have gained significant attention due to their tunable optical properties, thermal stability, and broad applicability in disciplines such as photothermal therapy, imaging, and biomedicine \cite{intro_HUANG201013,Liz-Marzán_intro,Dreaden2012,Richardson2006_intro,BRAVO2024_intro,Qin2012_intro}. Gold core silica shell nanoparticles have lately become a main focus of research due to their wide range of fundamental and technical applications \cite{Chen-goldsilica,Chen2-goldsilica,xie-goldsilica} such as cancer therapy \cite{hirsh2003_intro,day2024_intro} and energy conversion applications \cite{Neumann2013_intro,cui2023_intro}. The particular structure of gold core-shell nanoparticles allows for precise control of their localized surface plasmon resonance, which is highly sensitive to both the core size and the dielectric environment provided by the silica shell \cite{link_1999intro, maier2007_intro, Khlebtsov2010_intro}. In addition, the strong optical absorption and subsequent heat generation under laser irradiation make gold nanoparticles attractive for photothermal applications~\cite{Jain2007_intro, Baffou2009_intro, GOVOROV2007_intro}. Specifically, gold–silica core–shell systems stand out as ideal candidates for enhanced control and tunability~\cite{moularas2019,mertens2024,larquey2025}. 

It has been early recognized that coating a colloidal gold nanoparticle with a silica shell may facilitate heat transfer to the aqueous environment under pulse illumination~\cite{hu2003}. 
This effect has been proposed to explain the improved photoacoustic properties of gold-core silica-shell nanoparticles in water compared to their bare gold counterparts~\cite{Chen-goldsilica,Chen2-goldsilica,Au-Sio2_1_intro}. 
The relative enhancement displays a strong dependence on silica thickness, indicating a balance between improved interfacial thermal conduction and bulk thermal effects, highlighting the potential for silica coatings on thermal transfer around gold nanoparticles.
However, recent experimental and computational investigations demonstrated the deterioration of the photoacoustic response when the silica thickness increases~\cite{pang2020,Lombar_eph}. Alkurdi et al. 
showed by employing a two-temperature model that faster thermal transfer across gold core-silica shell nanoparticles stems from a strong coupling between the metal electrons and the dielectric silica shell~\cite{alikurdigoldsilica}. The existence of this channel is further supported by transient thermoreflectance measurements on gold films deposited on silicon and silica substrates, which reveal the existence of a direct energy exchange channel at metal-dielectric interfaces~\cite{e-phhopkins,Guo2012_eph}.
Importantly, this metal electron-silica energy pathway has also been identified as a key contributor to the enhanced photoacoustic response of core–shell nanoparticles~\cite{Lombar_eph,xie-goldsilica}.
In particular, Xie et al. show experimentally that a thin silica coating (thickness smaller than 6 nm) can enhance the photoacoustic signal by up to $400$\% compared to bare gold nanoparticles when irradiated with a picosecond pulsed laser.
This enhancement is interpreted by the strong out of equilibrium electron-phonon coupling at the gold-silica interface.

To model the ultrafast dynamics of electron and lattice temperatures in illuminated metallic nanostrutures, continuous approaches based on the two-temperature model have been traditionally employed~\cite{xie-goldsilica,Chen-goldsilica,moon_goldsilica,alikurdigoldsilica,Chen2-goldsilica}. However, such models generally neglect the atomistic features of heat transport and interfacial phenomena that are critical in nanoscale systems under pulse illumination~\cite{JulienPhysRevB.110.115437,gonzalez-colsa2023}. 

Here we go beyond these limitations by presenting an extension of the two-temperature model, that is coupled with molecular dynamics simulations, which offers an atomistic description of nanoscale heat transfer. The plasmonic response of the heterogeneous nanoparticles is modeled using an extension of Mie theory which takes into account electronic temperature corrections.

With this model in hand, we analyze the thermal response of illuminated gold core-silica shell nanoparticles immersed in water, with both dense and porous silica shells (with 50\% porosity). We consider three laser pulse durations (100\,fs, 10\,ps, and 1\,ns) and explore a range of laser fluences between 0.05 and 5\,mJ/cm$^2$. For all pulse durations, we observe a consistent trend: as the silica shell thickness increases, the time required to heat the surrounding water also increases, whether or not an interfacial electron-phonon coupling is considered at the gold silica interface. However, for nanoparticles with thin dense silica shells (5\,nm), and laser pulses of 100\,fs and  10\,ps, the interfacial electron-phonon coupling between gold and silica promotes rapid thermal equilibration between the core and the shell. This energy transfer across the core-shell interface facilitates heat transfer to the surrounding water, thereby accelerating the heating of the water interface compared to bare gold nanoparticles. 
In the light of this extended two temperature model, we interpret the recent photoacoustic experiments by Xie et al \cite{xie-goldsilica}. 

The structure of this article is as follows. In Section~\ref{sec:methods}, we detail the theoretical and computational approaches employed in this work, including the two-temperature model, plasmonic absorption calculations, molecular dynamics simulations, and the generation of dense and porous amorphous silica structures. We also describe the methodology considered to evaluate interfacial thermal conductance in gold–silica–water systems. Section~\ref{sec:results} presents and discusses the main results, focusing on the optical absorption characteristics of core–shell nanoparticles, the interfacial thermal conductance extracted from atomistic simulations, and the transient thermal response of the systems under laser excitation. Finally, Section~\ref{sec:conclusion} provides concluding remarks and perspectives for future work.

\begin{figure*}
    \centering
    \includegraphics[width=1\linewidth]{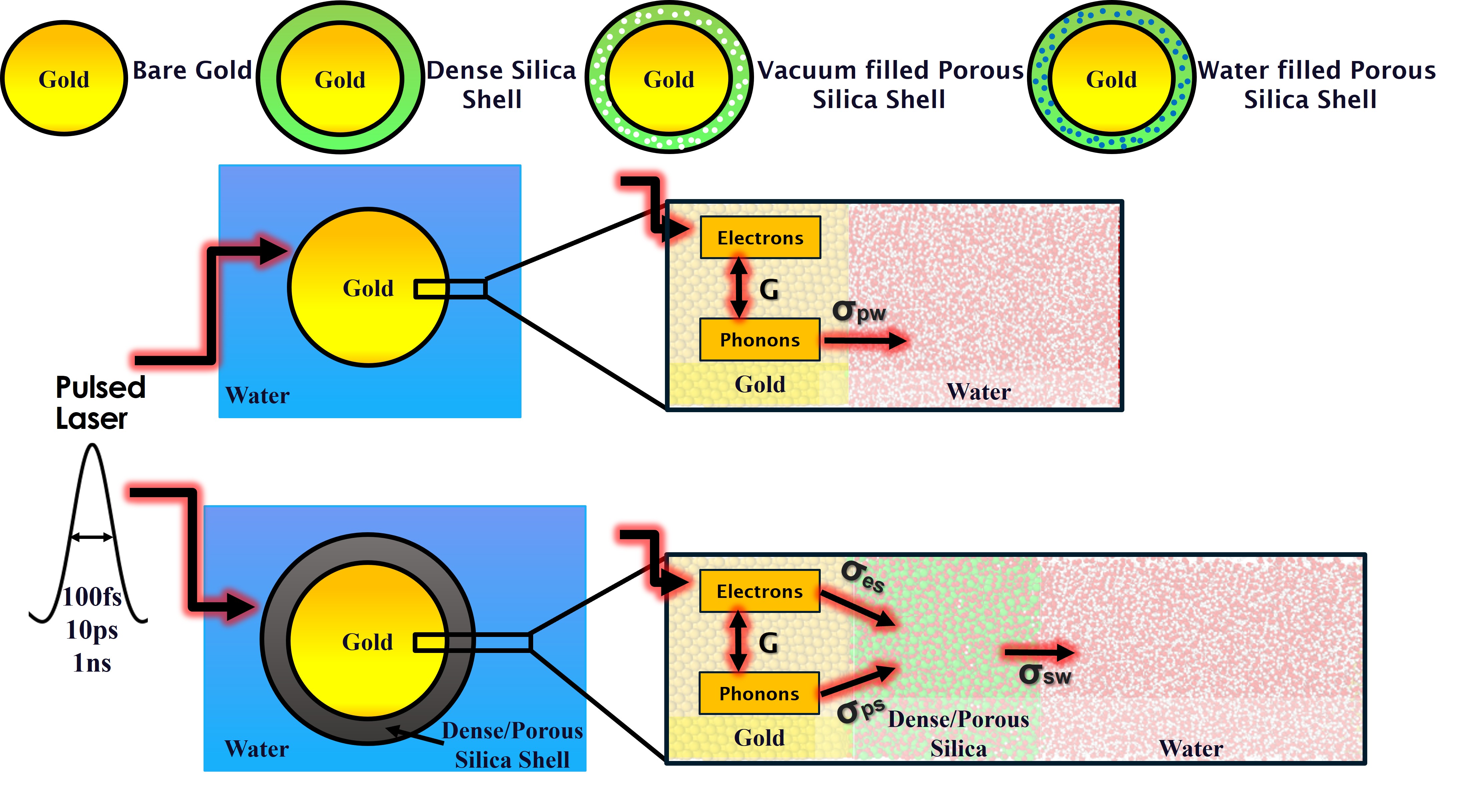}
    \caption{Illustration of the different energy channels from a bare gold and gold core-dense or porous silica shell nanoparticle heated by a pulsed laser  of 100fs, 10ps and 1ns laser pulse. \(G\) is the coupling constant between electrons and phonons in the metal. \(
    \sigma_{\mathrm{pw}},
    \sigma_{\mathrm{es}}, \sigma_{\mathrm{ps}}, \sigma_{\mathrm{sw}}
    \) are the interfacial conductance at the gold phonon water, gold electron silica, gold phonon silica, and silica phonon water interfaces, respectively.}
    \label{fig:NPinwater}
\end{figure*} 
\section{Theory and Methods} \label{sec:methods}
Our model combines the two-temperature model (TTM) with molecular dynamics (MD) simulations.  Special attention is placed on precisely calculating the nanoparticle's absorbance efficiency by integrating electronic temperature corrections to the refractive index of gold and using the Bruggeman approximation to characterize the complex optical features of the composites.

\subsection{Two Temperature Model} \label{sec:th_ttm}
The system under investigation is a core shell nanoparticle composed of a gold core surrounded by a silica shell, immersed in water and subjected to pulsed laser illumination as shown in Fig. \ref{fig:NPinwater}. Upon laser excitation, the free electrons in the metallic core absorb the light energy almost instantaneously, leading to a rapid rise in their temperature. This excess energy is then transferred to the shell via two mechanisms: directly, through electron–phonon thermal conductance at the core–shell interface (characterized by \(\sigma_{es}\)), and indirectly, through electron–phonon coupling within the core (governed by the coupling factor \(G\)) followed by phonon–phonon thermal coupling at the interface (characterized by \(\sigma_{ps}\)). As heat diffuses through the shell, a thermal flux proportional to the silica–water conductance \(\sigma_{sw}\) is generated at the shell–water interface, facilitating energy dissipation into the surrounding water. 

For comparison, a bare gold nanoparticle is used as a reference system throughout this study as shown in Fig. \ref{fig:NPinwater}. In this case, simultaneously, electrons and phonons within the metallic core reach thermal equilibrium, producing a thermal flux at the metal–water interface mediated by the interfacial phonon–phonon conductance \(\sigma_{pw}\). 

All these thermal processes are described by the following equations that are inspired by the TTM model~\cite{1974_TTM}:
\begin{align}
    V_e c_e \frac{\partial T_e(t)}{\partial t} &= -V_e G \left(T_e - T_p\right) - S_e \sigma_{es} \left[T_e - T_s(R_c,t)\right] \nonumber \\
    &\quad + P(t), \label{eq1} \\
    V_p c_p \frac{\partial T_p(t)}{\partial t} &= V_e G \left(T_e - T_p\right) - S_c \sigma_{ps} \left[T_p - T_s(R_c,t)\right], \label{eq2} \\
    c_s \frac{\partial T_s(r,t)}{\partial t} &= k_s \nabla^2 T_s, \label{eq3} \\
    c_w \frac{\partial T_w(r,t)}{\partial t} &= k_w \nabla^2 T_w, \label{eq4}
\end{align}

Here, the subscripts denote the component: \(e\) for the electrons in the gold core, \(p\) for the phonons in the gold core, \(s\) for the silica shell, and \(w\) for the surrounding water. \(T\), \(c\), and \(k\) represent respectively the temperature, heat capacity, and thermal conductivity of the corresponding component.

The parameters \( R_c, S_c, \) and \( V_c \) denote the core radius, surface area, and volume, while \( R = R_c + d \) represents the total nanoparticle  radius, with \( d \) being the shell thickness. The shell volume is given by \( V_s = V - V_c \).

The energy absorbed by the electrons, \( P(t) \), follows a Gaussian temporal profile \cite{wang_gaussian}:

\begin{equation}
    P(t) = \frac{2\sqrt{\ln2}}{\sqrt{\pi} \tau_p} F C_{\text{abs}} \exp\left(-\frac{4\ln2}{\tau_p^2} (t - 1.5\tau_p)^2\right), 
\end{equation}

where \( \tau_p \) is the pulse duration, and \( F \) is the laser fluence. 

The absorption cross-section, $C_{\text{abs}}$, depends on the temperature of the metal core, the dielectric shell, and the nanoparticle radii and will be given by an extended Mie theory that will be detailed in the next subsection. Here the thermophysical properties of the core-shell nanoparticle \( \sigma, k \) are obtained from MD simulations, ensuring accurate coupling between TTM and an atomistic level description.

The continuity of the heat flux across the interfaces results in the following boundary conditions at the core shell interface (\( r = R_c \)) and the shell water interface (\( r = R \)):

\begin{align}
-k_s \nabla T_s(R_c,t) \cdot \mathbf{n} 
&= \sigma_{es} \left[T_e - T_s(R_c,t)\right] \nonumber \\
&\quad + \sigma_{ps} \left[T_p - T_s(R_c,t)\right], \label{eq:interface_core}
\end{align}
\begin{align}
-k_s \nabla T_s(R,t) \cdot \mathbf{n} 
&= \sigma_{sw} \left[T_s(R,t) - T_w(R,t)\right] \nonumber \\
&= -k_w \nabla T_w(R,t) \cdot \mathbf{n}. \label{eq:interface_water}
\end{align}

where $\mathbf{n}$ is the unit vector normal to the interface and pointing outwards.
Last, as regards the initial conditions, all components are assumed to be at the uniform room temperature \( T_0 = 300\) K. 
\begin{table}[h!]
    \centering
    \resizebox{\columnwidth}{!}{%
    \footnotesize
    \begin{tabular}{lcccccc}
        \toprule
        \textbf{Quantity} & \textbf{Electron} & \textbf{Phonon} & \textbf{Dense Silica} & \textbf{Porous Silica} & \textbf{Water} & \textbf{Unit} \\
        \midrule
        \( k \) & 320.0 & 3.0 & 0.98 & 0.57 & 0.78 & \( \text{W m}^{-1}\text{K}^{-1} \) \\
        \( c \) & 0.0197 & 2.35 & 1.01 & 1.01 & 4.11 & \( \text{MJ m}^{-3}\text{K}^{-1} \) \\
        \bottomrule
    \end{tabular}%
    }
    \caption{Thermophysical parameters at \( T_0 = 300 \, \text{K} \) for the different system components, including dense and porous silica: thermal conductivity \(k\) and heat capacity \(c\).}
    \label{tab:thermophysical}
\end{table}

The thermophysical parameters relevant to gold core nanoparticles with dense or porous silica shells are listed in Table~\ref{tab:thermophysical}. The properties of bulk water, dense and porous silica are obtained from molecular dynamics simulations, while the electron and phonon parameters for gold are taken from Ref.~\cite{alikurdigoldsilica}. Additional parameters related to interfacial thermal conductance and thin film thermal conductivities are presented in Table~\ref{tab:itc_k_silica} in section~\ref{sec:results}. All properties are taken at \( T_0 = 300 \, \text{K} \), with two quantities requiring special consideration for their temperature dependence. The first is the electronic heat capacity, \( c_e = \gamma T_e \), where \( \gamma \) is the Sommerfeld constant of the metal. For gold, \( \gamma = 65.6 \, \text{J m}^{-3} \text{K}^{-2} \) \cite[]{Kittel86}. The second is the interfacial electron phonon conductance at the gold core-silica shell interface, which can be expressed as:
\begin{equation} \label{eq:e-ph}
    \sigma_{es}(T_e) = A + B T_e
\end{equation}
with the constants \( A = 96.1 \, \text{MW m}^{-2} \text{K}^{-1} \) and \( B = 0.18 \, \text{MW m}^{-2} \text{K}^{-2} \) \cite{Hopkins-thermalcond,alikurdigoldsilica,lombard_jpcm2015}.

\subsection{Plasmonic Calculations} \label{sec:plsmonic}
In this study, we use light scattering theory for coated spheres to investigate the optical absorption properties of core-shell nanoparticles. Mie theory is widely employed  to calculate electric field distributions as well as absorption and scattering efficiency factors of spherical particles \cite{Tuersun:13,Mundy:74,Fu:01}. Aden and Kerker were the first to derive an analytical solution for the electromagnetic fields and scattering parameters of a coated sphere \cite{AdenKerker10.1063/1.1699834}. Later, exact solutions for spherical particles were developed \cite{Kai:94,Mackowski:90}. The absorption efficiency $Q_{\text{abs}}$ is defined as the ratio of the absorption cross section to the geometrical cross section, which corresponds to the projected area of the particle on a plane perpendicular to the incident light. For a coated sphere, which has two radii $r_1$ and $r_2$, where $r_1$ represents the core radius and $r_2$ is the total radius including the shell, the absorption efficiency at a given wavelength $\lambda$ can be expressed by an infinite series \cite{CraigF.BohrenDonaldR.Huffman}:

\begin{equation}
Q_{\text{abs}} = \frac{2}{x_2^2} \sum_{n=1}^{\infty} (2n + 1) \left[ \text{Re}(a_n + b_n) - |a_n|^2 - |b_n|^2 \right]
\end{equation}

Here, \( x_2 = {2 \pi r_2 n_m}/{\lambda} \), is the geometric parameter of the shell, where \( n_m \) is the refractive index of the surrounding medium, which is water in this work. The scattering coefficients \( a_n \) and \( b_n \) depend on the optical constants and the refractive index of the core (\( n_1 \)) and the shell (\(n_2 \)) and can be calculated using the expressions detailed in the Supplemental Material.

Due to the fact that conduction electrons in gold heat up by absorbing the instantaneous (kinetic) energy from a laser (thereby increasing their temperature), one must account for the elevated electronic temperature when calculating the refractive index of the core. We model a temperature dependent dielectric function for gold, \(\varepsilon_{\mathrm{total}}(\omega,T_e)\), by superimposing a Drude-based intraband component, 
\(\displaystyle \varepsilon_{\mathrm{intra}}(\omega,T_e) = \varepsilon_{\infty} - \frac{\omega_p^2(T_e)}{\omega^2 + i\,\omega\,\nu(T_e)}\), and a Lorentz-based interband component, 
\(\displaystyle \varepsilon_{\mathrm{inter}}(\omega,T_e) = \sum_{j=1}^5 \frac{a_j(T_e)\,}{\omega_j^2(T_e) - \omega^2 - i\,\omega\,\Gamma_j(T_e)}\) \cite{Drudemodel,DrudeLorentzmodel}. The Drude-Lorentz framework is widely employed for modeling the optical response of various materials, as it naturally incorporates both free carrier (intraband) and interband transitions \cite{DL-example1,DL-example2,DL-example3,DL-Ndione}.
Here, \(\omega\) is the electromagnetic angular frequency, \(T_e\) is the electron temperature, and \(\varepsilon_{\infty}\) is the high-frequency dielectric constant accounting for the polarization of core electrons. The plasma frequency \(\omega_p(T_e)\) depends on the electron charge \(e\), vacuum permittivity \(\epsilon_0\), effective mass \(m_{sp}^*\) (approximated here by the free electron mass), and the temperature dependent conduction electron density \(n_{sp}(T_e)\) via 
\(\displaystyle \omega_p^2(T_e) = \frac{e^2\,n_{sp}(T_e)}{\epsilon_0\,m_{sp}^*}\). 
In turn, \(n_{sp}(T_e)\) is found by integrating the product of the partial density of states of the sp band \(D_{sp}(E)\) and the Fermi–Dirac distribution \(f(E,T_e,\mu)\): 
\(\displaystyle n_{sp}(T_e) = \int \mathrm{d}E\, f(E, T_e, \mu)\,D_{sp}(E)\) \cite{Ndione-old}. 
The collision frequency \(\nu(T_e)\) similarly reflects enhanced scattering at high \(T_e\). For the interband (Lorentz) term, five oscillators of the form 
\(\displaystyle \frac{a_j(T_e)\,}{\omega_j^2(T_e) - \omega^2 - i\,\omega\,\Gamma_j(T_e)}\) 
capture transitions from the \(d\)-band to the Fermi level \cite{JohnsonandChristy,DL-Ndione}. The five resonance frequencies used here were
taken from \cite{DL-Ndione,DL-example1}. The amplitude \(a_j(T_e)\) is scaled from the 300\,K value, \(a_{j,0}\), by the ratio of Fermi–Dirac occupation factors at energy \(E_j\), 
\(\displaystyle a_j(T_e) = a_{j,0} \,\frac{f(E_j,T_e)}{f(E_j,300\,\mathrm{K})}\), 
where 
\(
f(E,T) 
\;=\;
\frac{1}{1 + \exp\!\bigl(\tfrac{E - \mu}{k_B\,T}\bigr)}
\)
is the Fermi–Dirac distribution, and \(E_j\) is a reference transition energy (taken as -\(\hbar\omega_{j,0}\)). In these expressions, \(\mu\) is the chemical potential of the metal, \(k_B\) the Boltzmann constant. We set the chemical potential \(\mu \) equal to the Fermi level of gold because our calculations indicate that its shift remains under 1\% even at an electronic temperature of 10000 K as shown in the Supplemental Material. The resonance and the damping parameter are defined respectively as, 
\(\displaystyle \omega_j(T_e) = \omega_{j,0} + \bigl[\chi(T_e) - \chi(300\,\mathrm{K})\bigr]\) 
and 
\(\displaystyle \Gamma_j(T_e) = \Gamma_{j,\mathrm{ei}}(300\,\mathrm{K}) + \Gamma_{\mathrm{ee}}(T_e)\) where  \(\chi(T_e) = \mu_0 - c\,k_B T_e\) and \(\Gamma_{j,\mathrm{ei}}(300\,\mathrm{K})\) is the electron–phonon broadening at 300\,K, while \(\Gamma_{\mathrm{ee}}(T_e)\) is the electron–electron scattering contribution \cite{Damping_Fourment}. 

Finally, the total dielectric function, 
\[\displaystyle \varepsilon_{\mathrm{total}}(\omega,T_e) = \varepsilon_{\mathrm{intra}}(\omega,T_e) + \varepsilon_{\mathrm{inter}}(\omega,T_e)\]

matches well experimental data at 300\,K while capturing hot electron effects at elevated \(T_e\).

For the amorphous silica shell, we adopt the dispersion relation for the refractive index of fused silica determined over the wavelength range of 0.21 to 3.71\,\(\mu\)m \cite{MalitsonnSiO2:65}:
\begin{equation} \label{silica_eq_abs}
    n^2_{\text{SiO}_2} - 1 = \frac{A\, \lambda^2}{\lambda^2 - {A'}^2} + \frac{B\, \lambda^2}{\lambda^2 - {B'}^2} + \frac{C\, \lambda^2}{\lambda^2 - {C'}^2}
\end{equation}
where \(\lambda\) is the vacuum wavelength in \(\mu\)m.  Similarly, the refractive index of water as a function of \(\lambda\) is given by \cite{Daimon:nh2O07}:
\begin{equation} \label{water_eq_abs}
n^2_{\text{H}_2\text{O}} - 1 = \frac{D\, \lambda^2}{\lambda^2 - E} + \frac{F\, \lambda^2}{\lambda^2 - G} + \frac{H\, \lambda^2}{\lambda^2 - I} + \frac{J\, \lambda^2}{\lambda^2 - K}
\end{equation}
The numerical values of the constants are summarized in Table~\ref{tab:CombinedConstants}.
\begin{table}[h!]
    \centering
    \footnotesize
    \begin{tabular}{lccc}
        \toprule
        \textbf{Material} & \textbf{Constant} & \textbf{Value} & \textbf{Units} \\
        \midrule
        \multicolumn{4}{c}{\textbf{Fused Silica}} \\
        \midrule
         & \(A\)   & 0.6961663   & --- \\
         & \(A'\)  & 0.0684043   & \(\mu\text{m}\) \\
         & \(B\)   & 0.4079426   & --- \\
         & \(B'\)  & 0.1162414   & \(\mu\text{m}\) \\
         & \(C\)   & 0.8974794   & --- \\
         & \(C'\)  & 9.896161    & \(\mu\text{m}\) \\
        \midrule
        \multicolumn{4}{c}{\textbf{Water}} \\
        \midrule
         & \(D\)   & 0.5684027565   & --- \\
         & \(E\)   & 0.005101829712  & \(\mu\text{m}^2\) \\
         & \(F\)   & 0.1726177391   & --- \\
         & \(G\)   & 0.01821153936   & \(\mu\text{m}^2\) \\
         & \(H\)   & 0.02086189578   & --- \\
         & \(I\)   & 0.02620722293   & \(\mu\text{m}^2\) \\
         & \(J\)   & 0.1130748688    & --- \\
         & \(K\)   & 10.69792721    & \(\mu\text{m}^2\) \\
        \bottomrule
    \end{tabular}
    \caption{Dispersion equation constants for fused silica and water, Eq. \ref{silica_eq_abs} and Eq. \ref{water_eq_abs}, respectively.}
    \label{tab:CombinedConstants}
\end{table}

To estimate the refractive indices of porous silica, we use the Bruggeman effective medium approximation (EMA) \cite{Bruggeman,LandEAM}, which models the effective refractive index of a composite material based on its porosity and the refractive indices of its constituents. The Bruggeman formula for a two-component medium is given by:

\begin{equation}
    p \frac{n_a^2 - n_{\text{eff}}^2}{n_a^2 + 2n_{\text{eff}}^2} + (1 - p) \frac{n_b^2 - n_{\text{eff}}^2}{n_b^2 + 2n_{\text{eff}}^2} = 0
\end{equation}

where \( p \) is the porosity, \( n_a \) and \( n_b \) are the refractive indices of the two components, and \( n_{\text{eff}} \) is the effective refractive index of the composite material. 

For porous silica filled with water, we set \( n_a = n_{\text{SiO}_2} \) (refractive index of fused silica) and \( n_b = n_{\text{H}_2\text{O}} \) (refractive index of water). When the pores are unfilled (vacuum), we assume \( n_b = 1.0 \). This method provides a reliable estimation of the optical properties of porous silica for different porosities and filling conditions (either with water or vacuum), enabling accurate modeling of the complex permittivity of a two-phase composite \cite{KhardaniEAM,SchmidtEAM,SelaEAM}.

However, when light interacts with small metallic nanoparticles, the particle size becomes smaller than the mean free path of free electrons.which is around 40 nm for gold at room temperature \cite{Gall_MFelec}. As a result, the dielectric function deviates from its bulk value due to increased electron surface collisions, which introduce an additional relaxation mechanism. Consequently, the dielectric function must be corrected to account for surface scattering, and it is given by \cite{Tuersun:13, kreibig1995optical}:

\begin{equation}
\begin{aligned}
\varepsilon(\omega, L_{\text{eff}}) =\,& \varepsilon_{\text{bulk}}(\omega) + \frac{\omega_p^2}{\omega^2 + i\omega\,\dfrac{v_f}{l_\infty}} \\
&- \frac{\omega_p^2}{\omega^2 + i\omega \left( \dfrac{v_f}{l_\infty} + \dfrac{A\,v_f}{L_{\text{eff}}} \right)}
\end{aligned}
\end{equation}

\begin{figure*}
    \centering
    \includegraphics[width=1\linewidth]{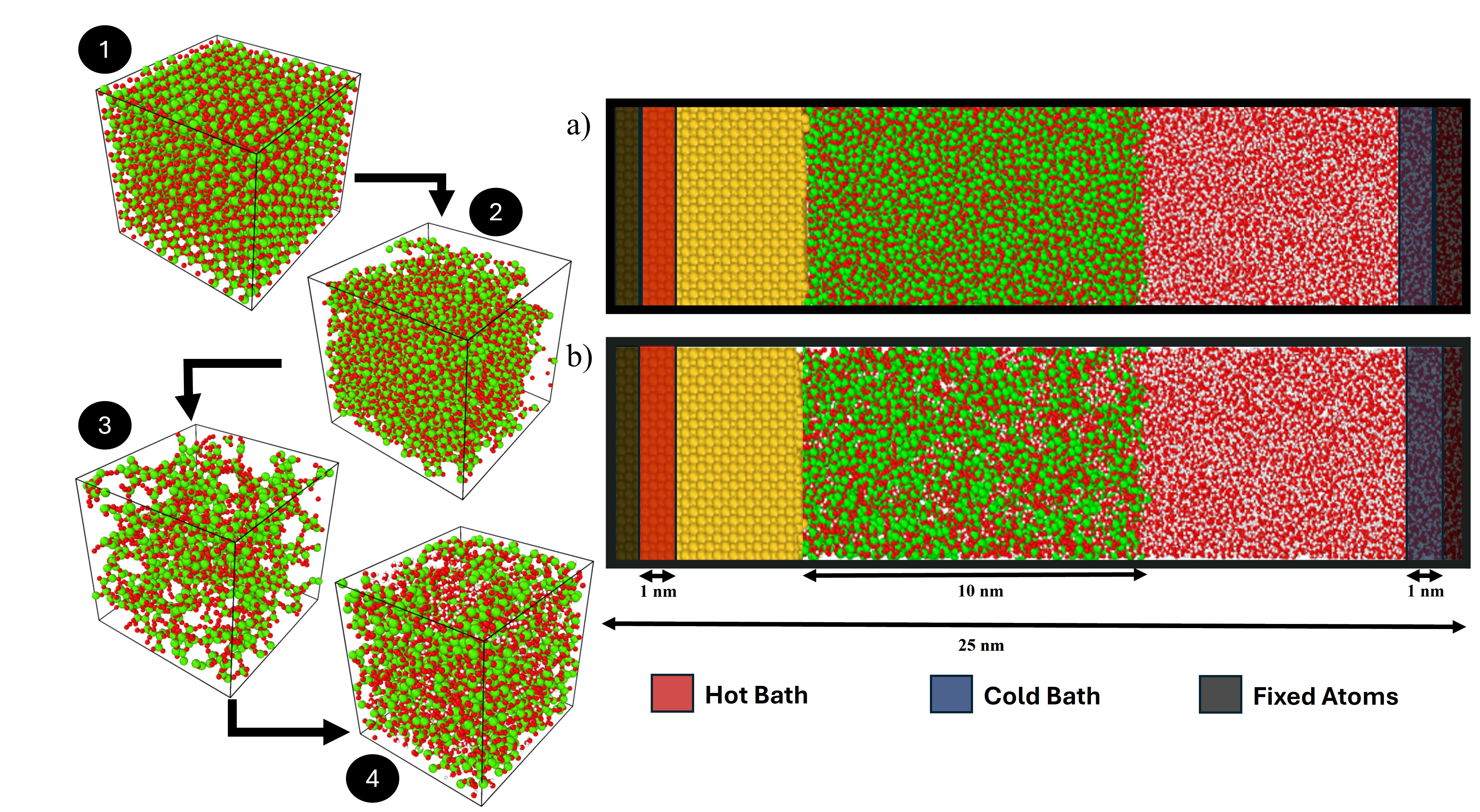}
    \caption{Illustration of the melt quench procedure used to generate porous silica from crystalline \(\alpha\)-cristobalite, comprising four steps. Step 1: The system is initialized in a crystalline state. Step 2: The crystalline structure is melted at 5000 K and the simulation box is gradually expanded to adjust the silica density. Step 3: The system is quenched by lowering the temperature gradually to 300 K, resulting in an amorphous, porous silica network. Step 4: Water molecules are introduced in the pores and the whole structure is equilibrated. After the structure is equilibrated, it is sandwiched in between gold and water for NEMD simulations. The schematic setup of the NEMD simulation is presented, where different atom types are indicated by distinct colors to illustrate the positions of frozen atoms and the thermal reservoirs in the sandwiched configuration, 
    shown for both: a) dense amorphous silica, and b) porous amorphous silica.}
    \label{fig:interfaces}
\end{figure*} 
Here, \( \varepsilon_{\text{bulk}} \) denotes the bulk dielectric function, \( \omega \) is the angular frequency of the incident light, \( \omega_p \) is the plasma frequency (\( \omega_p = 1.37 \times 10^{16} \,\mathrm{rad/s} \)), \( v_f \) represents the Fermi velocity of the metal (\( v_f = 1.4 \times 10^6 \,\mathrm{m/s} \)), \( l_\infty \) is the bulk mean free path of free electrons (\( l_\infty \approx 38 \,\mathrm{nm} \)), \( \varepsilon_\infty = 9.84 \) is the high-frequency dielectric constant, \( \gamma_{\text{bulk}} = 1.07 \times 10^{14} \,\mathrm{rad/s} \) is the bulk damping constant, \( A \) is a dimensionless parameter, and \( L_{\text{eff}} \) is the reduced effective mean free path of free electrons. For gold nanoparticles, we selected the parameter values from Refs.~\cite{HUANG201013, Rakic:98, Derkachovarevgold}.

\subsection{Molecular Dynamics Simulations}
To calculate the input parameters of the TTM, we perform MD simulations using the Large-scale Atomic/Molecular Massively Parallel Simulator (LAMMPS) \cite{LAMMPS}. The atomic trajectories are integrated using the velocity Verlet algorithm with a time step of 0.25 fs for all systems studied. The images of the simulated systems are generated using ovito software \cite{ovito}. 

For both dense and porous amorphous silica structures, the Tersoff potential function with Munetoh et al.’s parameters is employed to describe the interactions between silicon and oxygen atoms \cite{Tersoff-param}. A 12-6 Lennard-Jones potential is used to model the interactions between gold-gold, gold-silicon, and gold-oxygen atoms, with parameters specified by Heinz et al. for gold \cite{gold-heinz} and Rappe et al. for silicon and oxygen \cite{LJ-interfaces}.
The Heinz potential has been already employed in the modeling of colloidal gold nanoparticles~\cite{gutierrez-varela2022,rabani2023}.
The Lennard-Jones coefficients used at the interface between gold and silica are:
\(\varepsilon_{\text{Au-Si}} = 62.5 \, \text{meV}\) and \(\sigma_{\text{Au-Si}} = 3.39 \, \text{\AA}\) for gold-silicon interactions,
\(\varepsilon_{\text{Au-O}} = 42.9 \, \text{meV}\) and \(\sigma_{\text{Au-O}} = 3.056 \, \text{\AA}\) for gold-oxygen interactions. These parameters effectively reproduce the thermal conduction properties at the interfaces between the two materials as shown in \cite{JulienPhysRevB.110.115437}.

To model water, the TIP4P model is employed, and its functional form is presented in Eq.~(\ref{eq:water_potential}) \cite{waterTIP4P}. The TIP4P/2005 model has been shown to reproduce good thermal properties for water, accurately capturing the anomalous increase in thermal conductivity with temperature \cite{TC-water-TIP4P}. The interactions are governed by a combination of nonbonded and bonded potentials. The nonbonded interactions include the 12-6 Lennard-Jones potential, which accounts for repulsive and dispersive van der Waals forces, and the Coulomb potential, which captures the electrostatic interactions between the partial charges of the atoms. Coulombic interactions are computed using the particle-particle particle-mesh (PPPM) algorithm \cite{PPPM,PPPMdoi:10.1021/jp9518623}. These are described by the first two terms in Eq.~(\ref{eq:water_potential}). Bonded interactions include contributions from bond stretching and angle bending, which are expressed by the third and fourth terms in the following equation.

\begin{equation}
\begin{aligned}
    E_{\text{pot}} = &
    \sum_{i,j \, \text{(nonbonded)}} 
    4 \varepsilon_{ij} \left[ \left( \frac{\sigma_{ij}}{r_{ij}} \right)^{12} - \left( \frac{\sigma_{ij}}{r_{ij}} \right)^6 \right] \\
    & + \frac{1}{4 \pi \varepsilon_0} \sum_{i,j \, \text{(nonbonded)}} \frac{q_i q_j}{r_{ij}} \\
    & + \sum_{i,j \, \text{(bonded)}} 
    k_{r,ij} (r_{ij} - r_{0,ij})^2 \\
    & + \sum_{i,j,k \, \text{(bonded)}} 
    k_{\theta,ijk} (\theta_{ijk} - \theta_{0,ijk})^2,
\end{aligned}
\label{eq:water_potential}
\end{equation}

In this equation, \( r_{ij} \) represents the distance between atoms \( i \) and \( j \), while \( \varepsilon_{ij} \) and \( \sigma_{ij} \) are the Lennard-Jones parameters that describe the depth of the potential well and the finite distance at which the interatomic potential is zero, respectively.
For bond stretching, a harmonic potential is used where \(k_r = 26.03 \, \text{eV/\AA}^2\) is the bond force constant, and \(r_0 = 0.95 \, \text{\AA}\) is the equilibrium bond length between hydrogen (\(H_{\text{w}}\)) and oxygen (\(O_{\text{w}}\)) atoms. For angle bending, a harmonic potential is employed 
where \(k_\theta = 3.25 \, \text{eV/rad}^2\) is the angle force constant, and \(\theta_0 = 104.52^\circ\) is the equilibrium bond angle for the \(H_{\text{w}}-O_{\text{w}}-H_{\text{w}}\) angle. Using the mixing rule is a common method to obtain the nonbonding interaction potential parameters between different types of atoms \cite{goodmixing,TengFei_watersilica} , therefore we use the Lorentz-Berthelot rules \cite{lorentz} for determining the nonbonding interactions of Eq.~(\ref{eq:water_potential}) with parameters from \cite{water-amorphousislicaparam}.

The specific interaction parameters for water and its interactions with silica are summarized in Table~\ref{tab:water_silica_interactions}.

\begin{table}[h!]
    \centering

    \begin{tabular}{lccc}
        \toprule
        \textbf{Pair/Bond Type} & \(\varepsilon\) (meV) & \(\sigma\) (\AA) & \textbf{Charge} (\(q\)) \\
        \midrule
        H\(_{\text{w}}\)-H\(_{\text{w}}\) & 0.0 & 0.0 & \(+0.4238\) \\
        O\(_{\text{w}}\)-O\(_{\text{w}}\) & 8.03 & 3.1589 & \(-0.8476\) \\
        H\(_{\text{w}}\)-O\(_{\text{w}}\) & 0.0 & 0.0 & --- \\
        O\(_{\text{w}}\)-Si & 5.69 & 3.429 & --- \\
        O\(_{\text{w}}\)-O\(_{\text{silica}}\) & 6.51 & 3.124 & --- \\
        \bottomrule
    \end{tabular}
    \caption{Interaction parameters and atomic charges for TIP4P water and water-silica interactions in LAMMPS metal units.}
    \label{tab:water_silica_interactions}
\end{table}

\subsection*{Structural Generation of Dense and Porous Amorphous Silica}
\label{denseandporoussect}
To generate amorphous silica, we begin with the crystalline silica structure. Specifically, we use the \(\alpha\)-cristobalite phase, which has a space group symmetry of \(P4_1212\) and cell dimensions \(a = 4.99 \, \text{\AA}\) and \(c = 6.93 \, \text{\AA}\). The amorphous structures are created using the melt-quench technique \cite{France-Lanord_2014,JulienPhysRevB.110.115437}, following a multi-step process for porous silica \cite{VashishtaPhysRevLett.71.85,PATIL20212981}: starting from the crystalline \(\alpha\)-cristobalite structure at 300 K, the system is first equilibrated in the canonical ensemble (NVT) for stability as shown in step 1 of Fig. \ref{fig:interfaces}. Subsequently, the structure is heated to 5000 K over 1 ns, ensuring that the system enters the liquid phase and loses any memory of its initial configuration. The system is then held at this high temperature for an additional 1 ns to fully randomize its atomic positions. Finally, the temperature is decreased back to 300 K at a controlled rate of \(5 \, \text{K/ps}\), allowing the system to quench into an amorphous state. 

For porous silica, the same melt-quench procedure is adapted with modifications to achieve the desired porosity. During the melting phase at 5000 K, the simulation box dimensions are incrementally expanded, reducing the system density, as shown in step 2 of Fig. \ref{fig:interfaces}. This step is performed gradually to ensure a uniform distribution of voids within the material. The box is expanded until the target density is reached (50\% of the original crystalline silica density in this case). Once the desired porosity is achieved, the system is quenched back to 300 K at the same rate of \(5 \, \text{K/ps}\). The result is a stable, equilibrated porous silica structure with 50\% porosity as shown in step 3 of Fig. \ref{fig:interfaces}. This method ensures a uniform and controlled porosity distribution, allowing for reproducible material properties \cite{VashishtaPhysRevLett.71.85,PATIL20212981}. Both amorphous and porous silica structures are fully equilibrated at 300 K before further simulations are conducted. The thermal conductivity of the bulk amorphous silica is calculated using the Green-Kubo formalism \cite{GKPhysRevE.99.051301,boone2019heat}, yielding a value of \(0.98 \pm 0.09 \, \text{W/(m·K)}\) for dense amorphous silica which shows good agreement with values from other studies \cite[]{cond2silica,sio2cond_1}. The Radial Distribution Function (RDF) is computed, showing good agreement with previous simulation results in the literature \cite[]{asi-sio2conductivity} as shown in our previous study \cite{JulienPhysRevB.110.115437} and in the Supplemental Material. For porous silica with 50\% porosity, we find the thermal conductivity equal to \(0.25 \pm 0.02 \, \text{W/(m·K)}\) which also shows good agreement with previously reported values \cite{Yan.etal-porousislica,Costescu-thermalcond,Hopkins-thermalcond}. 
To prepare the porous silica structures filled with water, we begin with the pre-generated porous silica structure with 50\% porosity. This structure is sandwiched between two bulk water regions, allowing water molecules to diffuse into the pores over the course of the simulation. The system is equilibrated at room temperature, ensuring that all the pores are fully filled with water. Once the pores are completely filled, we remove the bulk water regions, leaving only the water molecules confined within the porous silica structure as shown in step 4 of Fig. \ref{fig:interfaces}. This process ensures a consistent and physically accurate representation of water confined within the pores of the silica structure. The final configuration is equilibrated to asses structural and thermodynamic stability before proceeding with further analyses. The thermal conductivity of the water-filled porous silica structure is \(0.57 \pm 0.03 \, \text{W/(m·K)}\), which is higher than that of the unfilled structure (due to the relative high thermal conductivity of water predicted by TIP4P/2005 model \cite{Ghanbarianwater-porousmedia}).

\subsection*{Interfacial Thermal Conductance and Thermal Transport in Gold-Silica-Water Systems}

The interfacial thermal conductance (ITC) in gold-silica-water systems is investigated using non-equilibrium molecular dynamics (NEMD) simulations by applying periodic boundary conditions along the x, y, and z directions. The Nose–Hoover equations of motion are used to integrate the atomic trajectories during all NEMD simulations \cite{nosee,HOOver}. The system consists of an amorphous silica thin film, dense or porous, sandwiched between gold and water, allowing for a detailed assessment of heat transport across both the solid-solid (gold - silica) and solid-liquid (silica - water) interfaces. The silica layer is considered as a thin film, and its thermal properties are evaluated based on its thickness. We investigate three different silica film thicknesses: 5 nm, 10 nm, and 20 nm. To enforce structural stability, the gold and water atoms at the system boundaries are fixed, while thermal baths are applied right next to the fixed atoms region to create a controlled temperature gradient as shown in Fig. \ref{fig:interfaces}. A steady state heat flux is established when the same amount of energy enters the system at the hot bath and exits at the cold bath, ensuring energy conservation as shown in Fig. \ref{fig:interfaces} for a 10 nm sandwiched amorphous silica thin film (Fig. \ref{fig:interfaces}a) and porous silica filled with water (Fig. \ref{fig:interfaces}b).

Each system has the same spatial dimensions: x = 5 nm, y = 5 nm, while z, the total system length depends on the silica thin film thickness. The water region contains approximately 7300 water molecules. The ITC at the gold-silica interface is determined by analyzing the steady-state temperature gradient across the system. Fixed atoms at the boundaries maintain structural integrity, while thermostats regulate heat flow across the interface. The heat flux (\(q\)) and the temperature jump (\(\Delta T_{\text{Au-SiO}_2}\)) at the interface between gold and silica are monitored, allowing the ITC to be computed as:

\begin{equation}
    G = \frac{q}{\Delta T_{\text{Au-SiO}_2}},
\end{equation}

where the heat flux is defined as:

\begin{equation} \label{eq:HF}
    q = \frac{\dot{E}}{A},
\end{equation}

with \(\dot{E}\) representing the rate of energy exchange in the system and \(A\) being the cross-sectional area of the interface. The steady-state temperature profile is averaged over 1 ns to ensure statistical accuracy. The computed ITC between gold and dense amorphous silica is consistent with our prior study \cite{JulienPhysRevB.110.115437} and other simulation results \cite{Raj-goldsilica} and is shown in Table \ref{tab:itc_k_silica}.

For the silica-water interface, the same NEMD approach is employed. The silica thin film serves as an intermediate medium for heat transfer between gold and water. The ITC at the silica-water interface is obtained by measuring the steady-state temperature drop (\(\Delta T_{\text{SiO2-H2O}}\)) at the silica-water interface. The interfacial thermal conductance is calculated as:

\begin{equation}
    G = \frac{q}{\Delta T_{\text{SiO2-H2O}}}
\end{equation}
The computed ITC is consistent with prior studies \cite{SUN2024} and also shown in Table \ref{tab:itc_k_silica}. 
The thermal conductivity (\(\kappa\)) of the silica thin film is calculated using the same heat flux equation (Eq.~\ref{eq:HF}) but normalized by the length (\(L\)) of the silica film, rather than the interface area:

\begin{equation}
    \kappa = \frac{q}{\Delta T},
\end{equation}

where  \(\Delta T\) is the average temperature jump across both interfaces. The calculated thermal conductivity are close enough to the ones calculated for bulk dense and porous silica in section \ref{denseandporoussect}. 
\begin{figure*}
    \centering
    \includegraphics[width=1\linewidth]{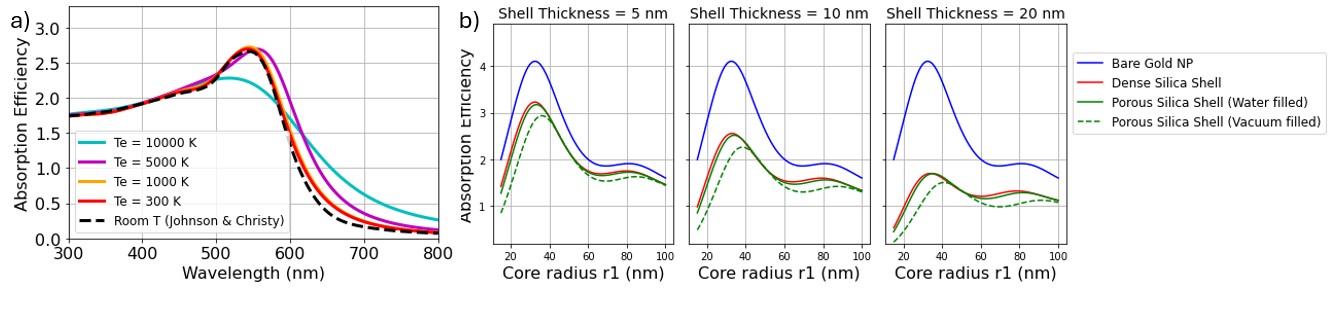}
    \caption{(a) Absorption efficiency for a bare gold nanoparticle in water (with electronic temperature corrections) as a function of wavelength, compared with experimental data from Johnson and Christy \cite{JohnsonandChristy} at 300\,K. (b) Absorption efficiency at a fixed wavelength of 532\,nm for a bare gold nanoparticle and for gold nanoparticles coated with three types of amorphous silica shells (dense silica, porous silica filled with vacuum, and porous silica filled with water) in water. The shell thickness is 5, 10, or 20\,nm, from left to right, and the results are plotted as a function of the gold nanoparticle core radius \(r_1\).}
    \label{fig:opt_changer1}
\end{figure*} 
\section{Results and Discussion}
\label{sec:results}

\subsection{Optical Absorption of Core Shell Nanoparticles}
We start by comparing our modeled gold refractive indices to the experimental results of Johnson and Christy \cite{JohnsonandChristy}. Fig. \ref{fig:opt_changer1}a) compares the absorption efficiency of a 50\,nm radius bare gold nanoparticle in water over the visible wavelength range at different electron temperatures (300\,K, 1000\,K, 5000\,K, and 10\,000\,K). The dashed black line corresponds to the room temperature experimental data, showing good agreement with our model at 300\,K. As the electron temperature increases, the absorption peak generally shifts to smaller wavelengths, showing how the optical response of gold changes with temperature. Interestingly, at \textit{T\textsubscript{e}} = 5000~K, the peak moves slightly to longer wavelengths compared to \textit{T\textsubscript{e}} = 1000~K. At even higher temperatures, \textit{T\textsubscript{e}} = 10000~K, the shorter wavelengths shift continues as expected. This trend stresses the necessity of properly accounting for temperature dependent characteristics in modelling of plasmonic heating.

In the following, we investigate the effect of shell thickness on the absorption for three specific thicknesses: 5, 10 and 20 nm. For the porous silica shell, we will focus on a single value of porosity equal to 50\%, as previously stated. To optimize the absorption, we fix the wavelength at 532 nm and change the radius \(r_1 \) of the gold nanoparticle. 
\begin{figure*}
    \centering
    \includegraphics[width=1\linewidth]{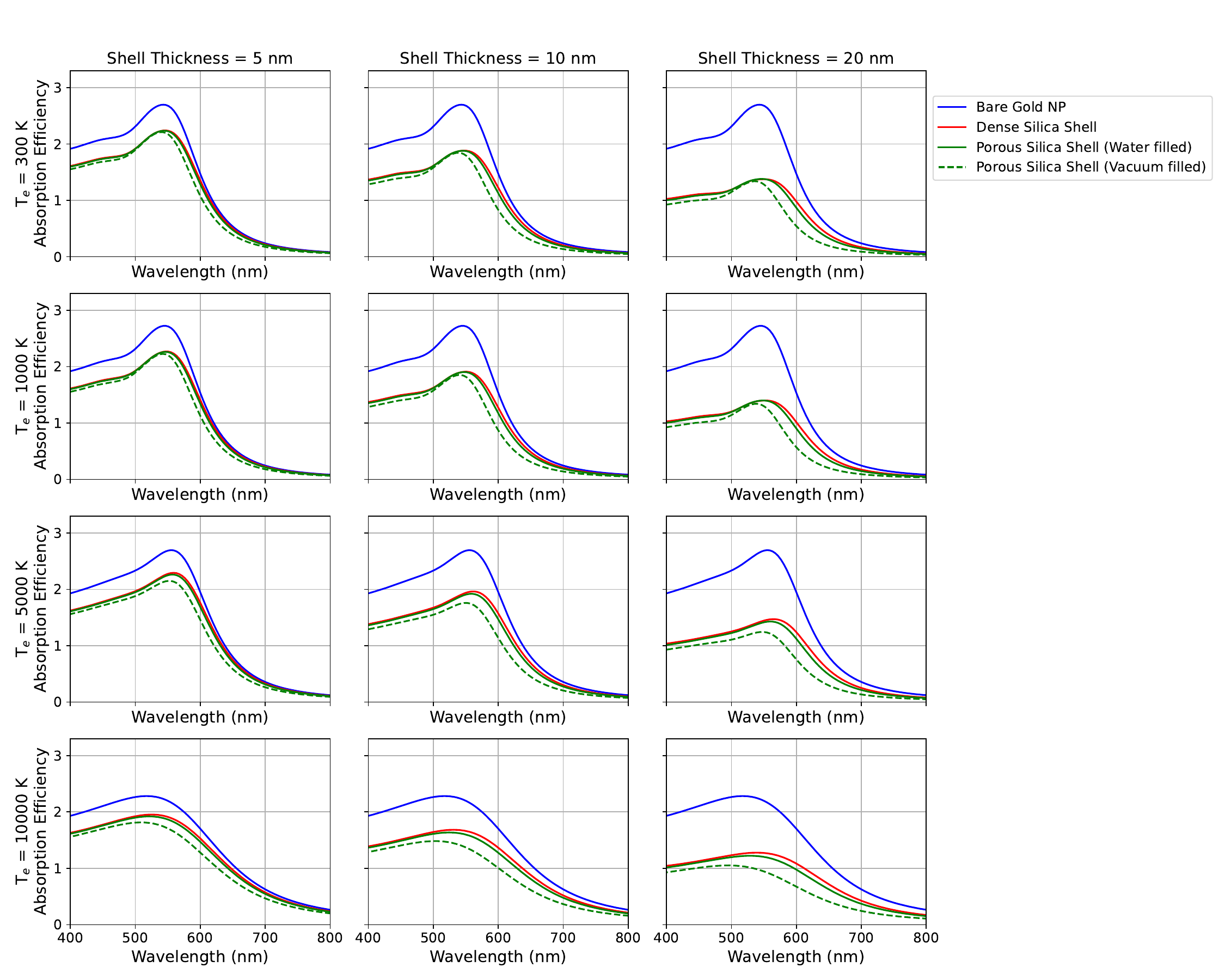}
    \caption{Absorption efficiency as a function of wavelength of bare and coated gold nanoparticles with amorphous silica shells and a fixed core radius of 50 nm. The figure is arranged in 3 rows and 4 columns, where the columns correspond to shell thicknesses of 5 nm, 10 nm, and 20 nm, respectively, and the rows correspond to electron temperatures of 300 K, 1000 K, 5000 K, and 10000 K.}
    \label{fig:opt_Tedep}
\end{figure*} 

Figure~\ref{fig:opt_changer1}b) shows the absorption efficiency at a fixed wavelength of 532 nm for a bare gold nanoparticle and for gold nanoparticles coated with three types of amorphous silica shells: dense silica, porous silica filled with vacuum, and porous silica filled with water. The results in Fig. \ref{fig:opt_changer1}b) show that for particles smaller than around 40 nm in radius, the absorption is essentially volume driven (Rayleigh regime), while for bigger particles, surface plasmon related effects dominate. As the radius exceeds ~40nm, the optical response of the nanoparticle depends more on its surface than on its volume. In all cases, bare gold nanoparticles exhibit significantly higher absorption efficiency compared to their core–shell counterparts. This difference is most pronounced at intermediate core radii (30–60 nm), where the plasmon resonance is strongest. Dense shells result in a higher absorption than porous ones, particularly for small radii, while water filled porous shells show a marginally higher absorbance efficiency than vacuum filled ones due to the higher dielectric contrast with gold.

Next, we fix the gold core radius at 50 nm and evaluate the effect of different silica shell configurations on the nanoparticle absorbance.
Fig.~\ref{fig:opt_Tedep} presents the absorption efficiency spectra for a 50 nm gold nanoparticle coated with amorphous silica shells, computed using the model presented in section \ref{sec:plsmonic}. The figure is arranged in 4 rows and 3 columns, with the rows corresponding to electron temperatures of 300, 1000, 5000, and 10\,000\,K, and the columns representing shell thicknesses of 5, 10, and 20\,nm, respectively (one must note that we do not reach an electronic temperature of 10,000 K in the simulations presented here).

Compared to the bare gold nanoparticle, all core–shell configurations exhibit a clear reduction in absorption efficiency, particularly in the visible range near the plasmon resonance. This decrease is most notable for thicker shells and vacuum filled porous structures, which dampen and red-shift the plasmon resonance due to the dielectric screening. For a given shell thickness, the absorption spectra for dense silica and water filled porous silica are nearly identical, suggesting that the effective refractive index of water filled porous silica is comparable to that of dense silica. In contrast, when the pores are filled with vacuum, the absorption efficiency is significantly lower, particularly near the plasmon resonance, because the reduced refractive index of the shell modifies the plasmon response. Moreover, as the electron temperature increases, the plasmon peak of the gold core narrows and eventually diminishes, reflecting a strong temperature driven modification of the resonance profile. Despite these changes, the relative differences among the three types of shell remain consistent across all electron temperatures, with vacuum filled porous silica consistently yielding a lower absorption efficiency in both the visible and near-infrared regions.

\subsection{Interfacial Thermal Conductance from Molecular Dynamics Simulations}
\begin{figure}[h]
    \centering
    \includegraphics[width=1\linewidth]{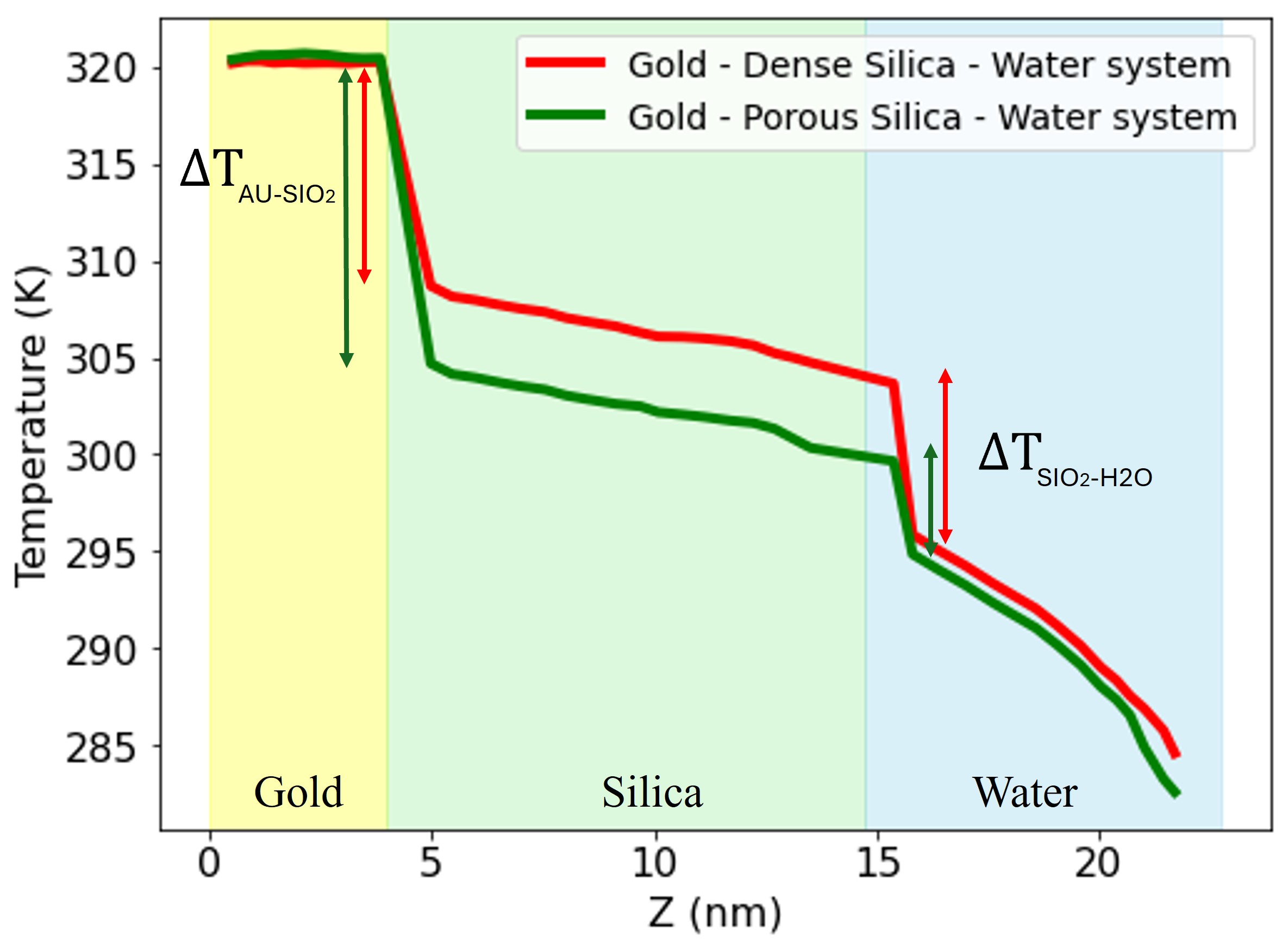}
    \caption{Temperature profile averaged from the NEMD simulations used to estimate the temperature jump at the gold amorphous silica and amorphous silica-water interfaces for both dense and porous amorphous silica.}
    \label{fig:temp_jump}
\end{figure}

\begin{table*}[ht]
\centering
\footnotesize
\begin{tabular}{lcccccc c}
\toprule
\multirow{2}{*}{\textbf{Shell Thickness}} & \multicolumn{2}{c}{\textbf{Gold–Silica ITC}} & \multicolumn{2}{c}{\textbf{Silica–Water ITC}} & \multicolumn{2}{c}{\textbf{Thermal Conductivity}} & \multirow{2}{*}{\begin{tabular}{c} \textbf{Units} \\ \textbf{Conductance/Conductivity} \end{tabular}} \\
\cmidrule(lr){2-3} \cmidrule(lr){4-5} \cmidrule(lr){6-7}  
 & Dense & Porous & Dense & Porous & Dense & Porous & \\
\midrule
5 nm  & $182\pm22$  & $137\pm11$ & $847\pm51$  & $1082\pm67$  & $0.95\pm0.08$ & $0.55\pm0.05$ & MW\,m$^{-2}$K$^{-1}$/W\,m$^{-1}$K$^{-1}$ \\
10 nm & $172\pm18$  & $125\pm9$  & $832\pm43$  & $1050\pm60$  & $0.98\pm0.09$ & $0.57\pm0.05$ & MW\,m$^{-2}$K$^{-1}$/W\,m$^{-1}$K$^{-1}$ \\
20 nm & $165\pm11$  & $115\pm8$  & $817\pm58$  & $1021\pm65$  & $1.00\pm0.06$ & $0.61\pm0.07$ & MW\,m$^{-2}$K$^{-1}$/W\,m$^{-1}$K$^{-1}$ \\
\bottomrule
\end{tabular}
\caption{Interfacial thermal conductance calculated with MD simulations at the gold–silica and silica–water interfaces and thermal conductivities of dense and porous silica for shell thicknesses of 5, 10, and 20 nm.}
\label{tab:itc_k_silica}
\end{table*}
The temperature profile presented in Fig. \ref{fig:temp_jump} provides valuable insights into the thermal conductance at the interfaces between gold and silica, as well as between silica and water. The observed temperature gradient across the gold-dense silica interface is relatively moderate, indicating efficient heat transfer and a lower thermal boundary resistance. In contrast, the gold-porous silica interface exhibits a steeper temperature drop, signifying a higher thermal resistance at the interface and lower interfacial thermal conductance. This difference arises from the porous structure, which introduces voids and reduces atomic scale contact, thereby impeding heat transport at the interface with gold.

\begin{figure*}
    \centering
    \includegraphics[width=1\linewidth]{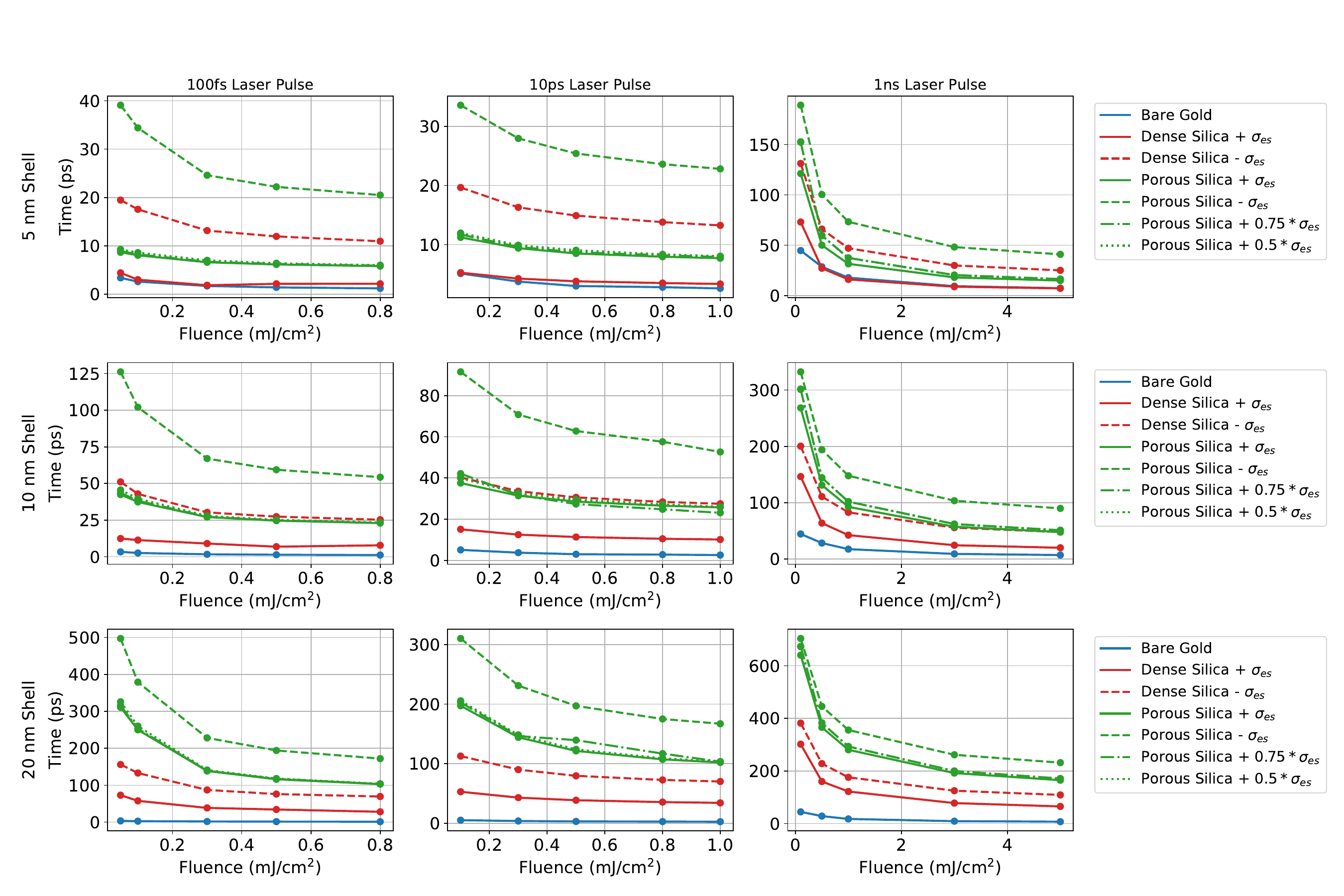}
    \caption{Influence of silica shell thickness and electron-phonon coupling on the nanoparticle response under laser pulses of 100\,fs, 10\,ps, and 1\,ns, aimed at achieving a 10\,K temperature rise at the nanoparticle water interface. In the figure, the columns correspond to different laser pulse durations (from left to right: 100\,fs, 10\,ps, and 1\,ns), while the rows correspond to different silica shell thicknesses (from top to bottom: 5, 10, and 20\,nm). Each subplot displays the time (ps) required to reach 10\,K temperature rise at the nanoparticle water interface as a function of laser fluence (mJ/cm$^2$). For comparison, the response of the bare gold nanoparticle is included.}
    \label{fig:timevschangeF_fspsns}
\end{figure*}

On the other hand, at the silica water interface, the trend is reversed. The temperature gradient between dense silica and water is steeper than that of porous silica and water, showing that the presence of pores enhances heat transfer at the silica-water interface. This can be attributed to the presence of nanoscale pores that facilitate energy exchange with water, likely due to enhanced phonon interactions and localized heat dissipation pathways \cite{nanoprousilicaexpTC}.
Overall, the analysis highlights a trade-off in the mechanisms of thermal transport: while porosity reduces thermal conductance at the gold-silica interface due to reduced contact, it enhances heat dissipation into water by improving interfacial coupling.

Table ~\ref{tab:itc_k_silica} summarizes the calculated interfacial thermal conductance at the gold–silica and silica–water interfaces, as well as the bulk thermal conductivities for both dense and porous silica shells, for shell thicknesses of 5, 10, and 20 nm. Notably, although minor variations are observed with changing shell thickness, the overall values remain of similar magnitude across the different thin film considered. This consistency indicates that, within the range of 5–20 nm, the interfacial and bulk thermal properties are largely determined by the intrinsic material characteristics rather than by the shell thickness.

\subsection{Transient Thermal Response of Core Shell Nanoparticles}

\begin{figure*}
    \centering
    \includegraphics[width=1\linewidth]{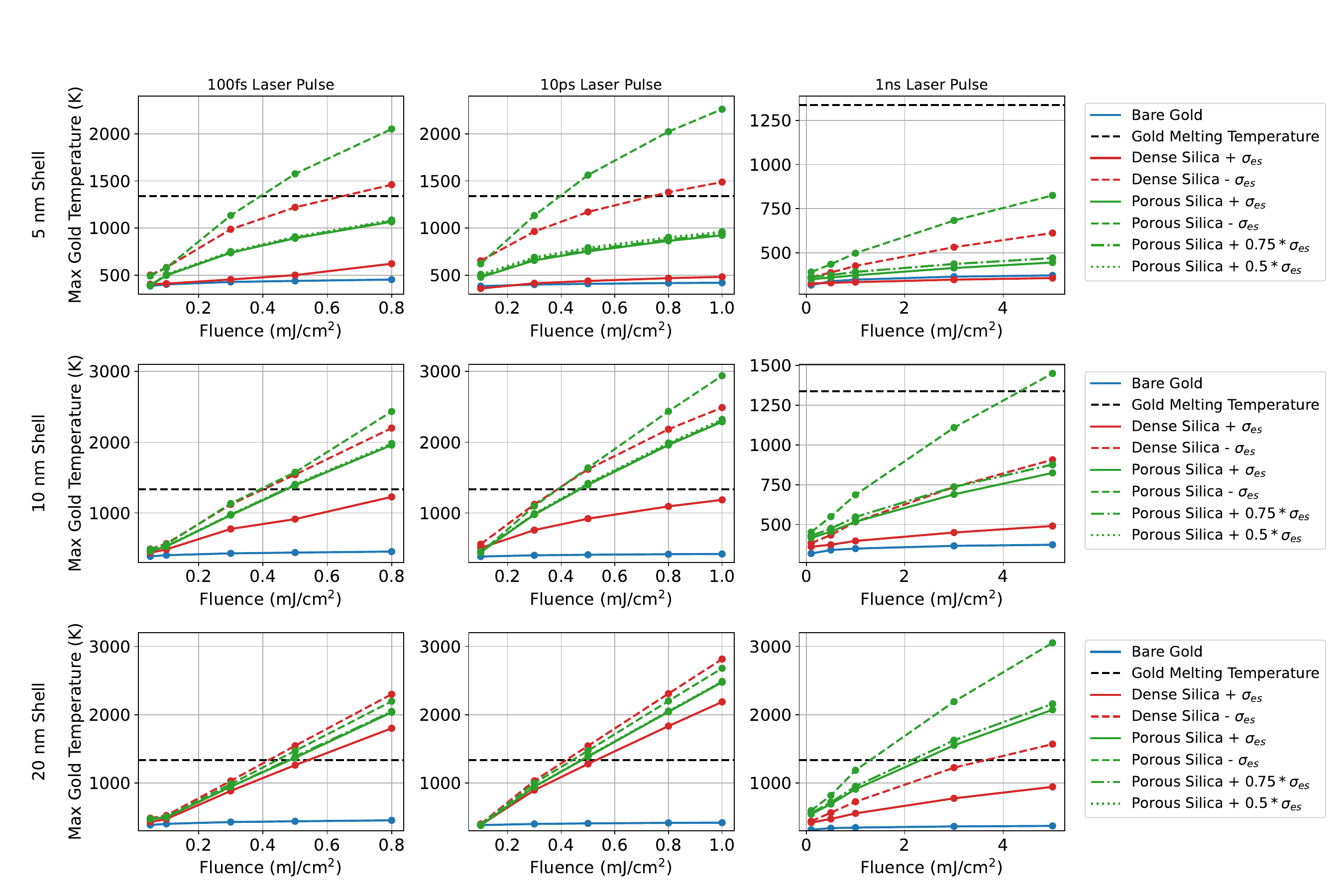}
    \caption{Influence of silica shell thickness and electron-phonon coupling on the maximum gold core temperature under laser pulses of 100\,fs, 10\,ps, and 1\,ns, aimed at achieving a 10\,K temperature rise at the nanoparticle water interface. In the figure, the columns correspond to different laser pulse durations (from left to right: 100\,fs, 10\,ps, and 1\,ns), while the rows correspond to different silica shell thicknesses (from top to bottom: 5, 10, and 20\,nm). Each subplot displays corresponding gold temperature (K) required to reach the nanoparticle response as a function of laser fluence (mJ/cm$^2$). Both the gold melting temperature and the response of bare gold nanoparticles are shown for comparison.}
    \label{fig:TgvschangeF_fspsns}
\end{figure*} 

First, we analyze the effect of laser fluence on the transient thermal response of gold-core nanoparticles coated with dense and porous silica shells, both with and without electron–phonon coupling at the gold–silica interface. For comparison, we also consider a bare gold nanoparticle. We explore laser pulse durations of 100fs, 10ps, and 1ns, while additional results for 1fs and 1ps pulses are presented in the Supplemental Material. They display close thermal responses with the 100fs and 10ps laser pulse durations, respectively. For the porous silica shells, with electron-phonon coupling between gold and silica, we introduce a dimensionless parameter $\zeta$ to account for the additional scattering induced by the pores. In our model, the effective electron phonon coupling between gold and porous silica is given by $\zeta\,\sigma_{es}$, and we consider three values for $\zeta$ (1, 0.75, and 0.5) to investigate its impact on the thermal dynamics.

Fig.~\ref{fig:timevschangeF_fspsns} and \ref{fig:TgvschangeF_fspsns} systematically illustrate the influence of laser fluence on the thermal response of gold core nanoparticles with 5\,nm, 10\,nm, and 20\,nm silica shells, considering both the presence and absence of electron-phonon coupling at the gold silica interface as previously mentioned. In these figures, the columns correspond to different laser pulse durations (from left to right: 100\,fs, 10\,ps, and 1\,ns), while the rows represent different silica shell thicknesses (from top to bottom: 5, 10, and 20\,nm). Fig.~\ref{fig:timevschangeF_fspsns} displays the time (ps) required to achieve a 10\,K temperature rise at the nanoparticle water interface as a function of laser fluence (mJ/cm$^2$), and Fig.~\ref{fig:TgvschangeF_fspsns} shows the maximum gold core temperature (K) with a horizontal black line indicating the melting temperature of gold. However, note that under the ultrafast pulse conditions considered here, the energy deposition is insufficient to melt the gold; achieving melting would require a longer pulse duration.

\begin{figure*}
    \centering
    \includegraphics[width=1\linewidth]{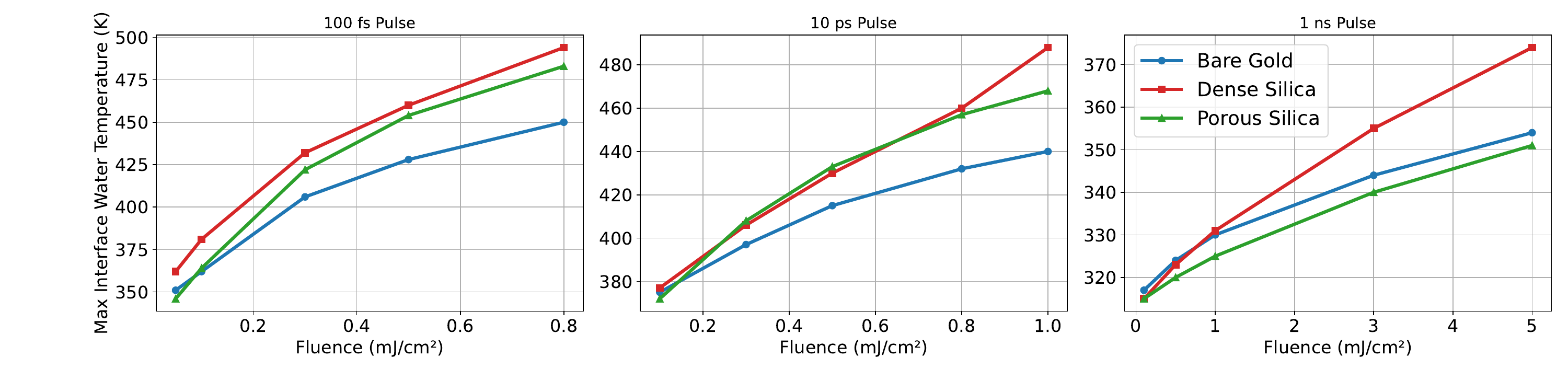}
    \caption{
    Evolution of the maximum interface water temperature at the nanoparticle water interface  under laser pulses of 100\,fs, 10\,ps, and 1\,ns (from left to right), for a bare gold, a 5\,nm dense silica, and a 5\,nm porous silica shells, including electron-phonon coupling for the silica shells as function of laser fluences \textit{F} in (mJ/cm$^2$).
}

    \label{fig:maxtemp}
\end{figure*} 

\begin{figure*}
    \centering
    \includegraphics[width=1\linewidth]{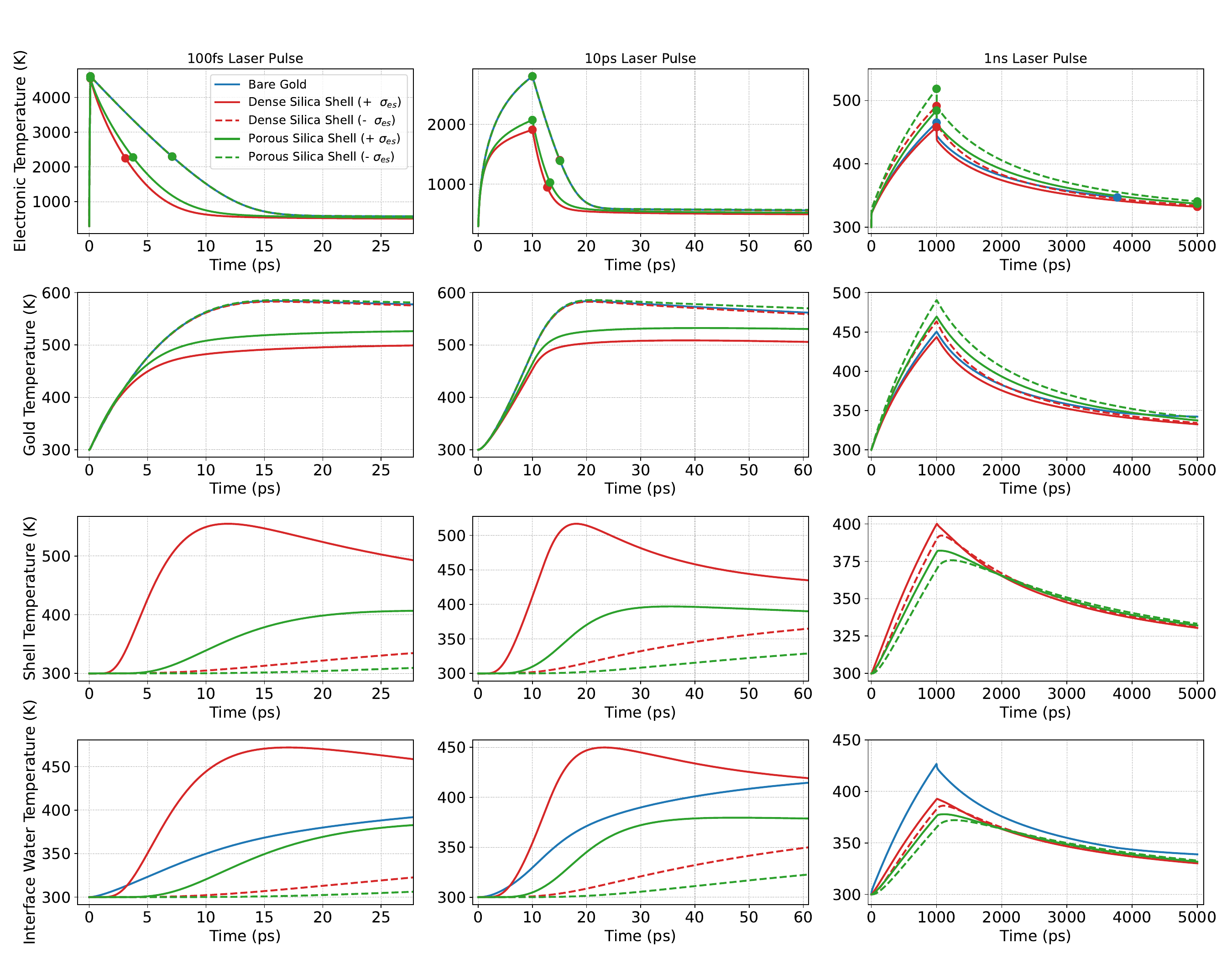}
   \caption{Time evolution of the electronic temperature, gold core temperature, silica shell temperature, and water temperature (measured at the nanoparticle water interface) for bare gold as well as gold–silica nanoparticles. Both dense and porous configurations are shown, with and without electron–phonon coupling (\(\sigma_{es}\)), under laser pulses of 100\,fs, 10\,ps, and 1\,ns. A constant shell thickness of 5\,nm and a fluence of 0.1\,mJ/cm\(^2\) is considered. Markers on the time evolution of the electronic temperature plots indicate the maximum and half-maximum points used to estimate the cooling rates via the half-maximum criterion. The simulations begin with an initial temperature of 300\,K, corresponding to thermal equilibrium prior to laser excitation. The cooling dynamics are monitored until the system temperature decreases back to 300\,K.}
    \label{fig:FixF_fspsns}
\end{figure*}

\noindent For a given fluence, the data indicate that shorter laser pulses (100\,fs and 10\,ps) produce a faster thermal response, with significantly shorter times to reach the specified temperature rise compared to the 1\,ns pulse. This behavior is attributable to the more impulsive energy delivery associated with ultrashort pulses, which results in higher peak electron temperatures and more abrupt energy transfer dynamics. Moreover, the presence of electron-phonon coupling (denoted by $+\sigma_{es}$) enhances the thermal response by facilitating efficient energy transfer from the gold core to the silica shell, thereby reducing the response time and the maximum gold temperature relative to cases where this coupling is absent ($-\sigma_{es}$). These effects are pronounced for all the considered shell thicknesses. For thin dense silica shells (5 nm), the enhanced electron–phonon coupling between the gold core and the silica shell enables a faster thermal response than that observed for bare gold nanoparticles. However, thicker shells (10 and 20 nm) tend to delay the response, and notably increase the heating time as well as the core temperature due to increased thermal resistance and the low thermal conductivity of the shell. In addition, our results show that with increasing laser fluence, the time required to reach the required 10\,K water interface temperature decreases, and the corresponding gold core temperature increases. However, caution must be taken when considering the high fluences and thicker silica shells used in our study, as the resulting temperatures may exceed the melting point of gold, potentially leading to nanoparticle damage. In the case of electron–phonon coupling between gold and porous silica, scaled by a factor \(\zeta\), the variations in the thermal response are minimal. As detailed in the model, the electron–phonon coupling is expressed in Eq.~\ref{eq:e-ph}. For ultrashort laser pulses (100\,fs), varying the coupling strength for porous silica by scaling factors (e.g., 1, 0.75, or 0.5) leads to only minor differences in the thermal response. This suggests that under such rapid excitation conditions, the precise magnitude of the coupling is not critical, and simply accounting for its presence already captures the dominant physical effects. However, as the pulse duration increases to 10\,ps and 1\,ns, the slower energy deposition allows the impact of different coupling strengths to become more pronounced, leading to observable differences in the cooling behavior and heat transfer dynamics.

Fig.~\ref{fig:maxtemp} presents the maximum temperature reached in our simulations at the nanoparticle water interface as a function of laser fluence under laser pulses of 100 fs, 10 ps, and 1 ns (from left to right). For all pulse durations, the maximum interface temperature increases monotonically with fluence. Across the different laser pulses, nanoparticles coated with a thin (5 nm) dense silica shell reach higher peak interface temperatures than those with a porous shell or bare gold. This indicates that the dense silica shell enhances heat transfer from the nanoparticle core to the surrounding water due to more efficient heat transfer at gold-dense silica interface.  The influence of pulse duration is also evident, shorter pulses (100 fs) lead to significantly higher temperature peaks for a given fluence, owing to rapid energy deposition and limited heat dissipation during the pulse.

Now, we fix the silica shell thickness at 5\,nm and the laser fluence at 0.1\,mJ/cm$^2$ and report the resulting thermal dynamics for the three nanoparticle configurations. At this value of the fluence, the energy deposited into the system is moderate, ensuring that the gold core remains well below its melting point and that the nanoparticles are far from the fragmentation threshold for the sizes considered \cite{kang2021frag}.

Fig.~\ref{fig:FixF_fspsns} displays the evolution of the electronic, gold core, silica shell, and water interface temperatures as a function of time for 100\,fs, 10\,ps, and 1\,ns laser pulses, respectively. In addition to the standard configurations where electron-phonon coupling between gold and silica is considered (denoted by “+ $\sigma_{es}$”), we also analyze cases in which this coupling is neglected (denoted by “– $\sigma_{es}$”), for both dense and porous silica shells. Upon laser excitation, the electronic temperature rises almost instantaneously, reaching a pronounced peak due to rapid energy absorption. Shortly thereafter, the gold core temperature increases as energy is transferred from the electrons via electron-phonon coupling, and the silica shell temperature gradually converges toward that of the gold core. Notably, when electron-phonon coupling at the gold silica interface is not taken into account, the temperature equilibration between the gold core and the silica shell is slower. 
\begin{figure*}
    \centering
    \includegraphics[width=1\linewidth]{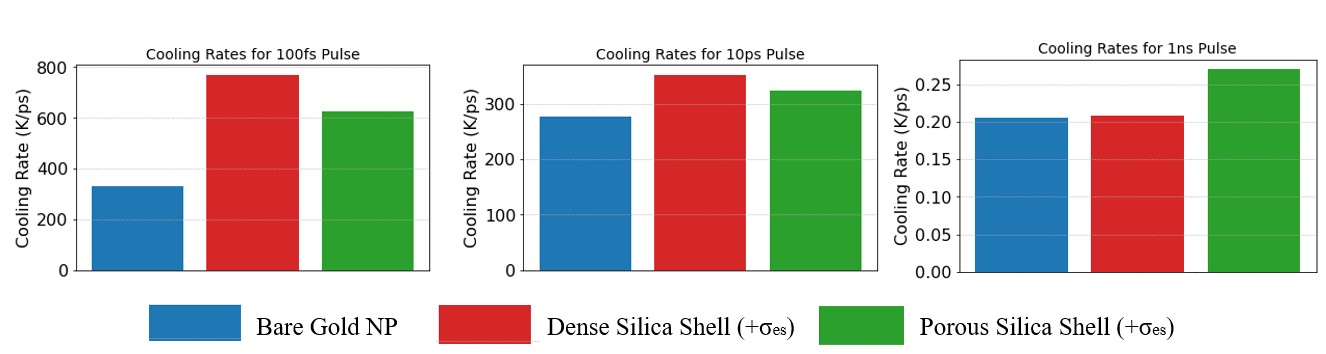}
   \caption{Histograms of the cooling rates for the different nanoparticle configurations under investigation: bare gold (blue) and silica shells (both dense and porous) with electron–phonon coupling (red and green, respectively), for laser pulse durations of 100fs, 10ps, and 1ns,  and a fixed fluence of 0.1\,mJ/cm\(^2\).}
    \label{fig:histocooling}
\end{figure*} 
\begin{figure*}
    \centering
    \includegraphics[width=1\linewidth]{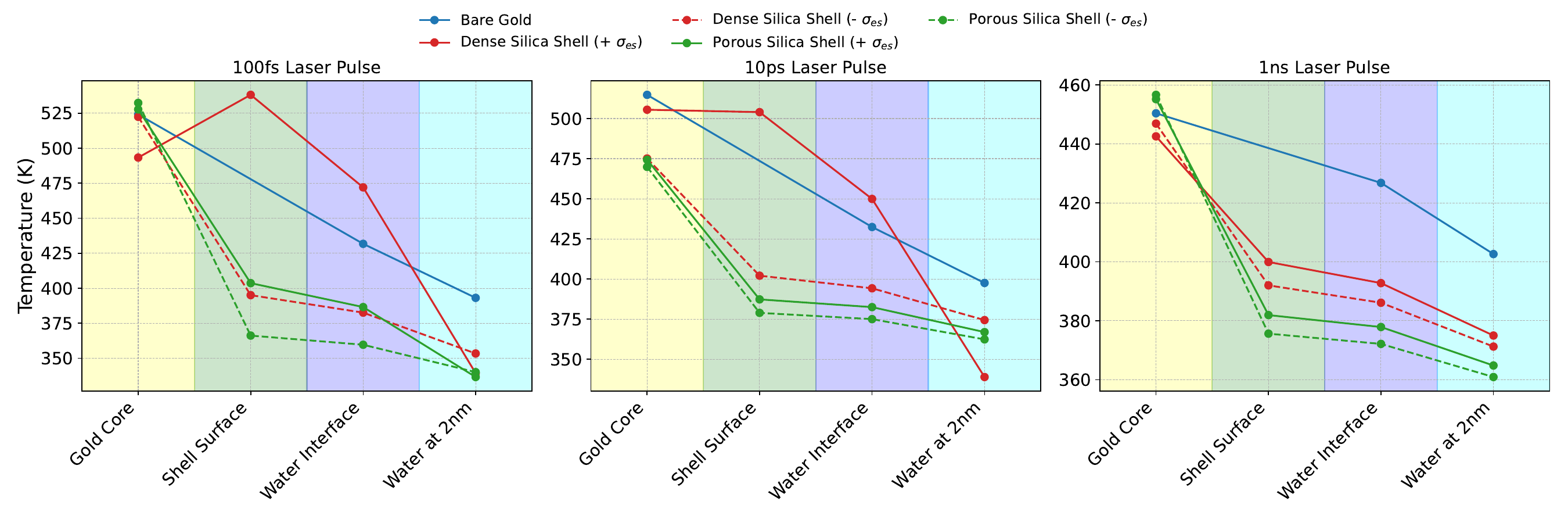}
    \caption{Temperature profiles of specific regions of the systems measured when the water temperature is maximum at the nanoparticle interface at fixed times for laser pulses of 100 fs, 10 ps, and 1 ns for a fixed fluence of 0.1\,mJ/cm\(^2\). In each panel, horizontal segments represent the measured temperatures at discrete regions of the system: gold core, silica shell (5 nm thick), water interface, and water at 2 nm  from the interface. The data compare the response of bare gold nanoparticles with that of nanoparticles coated with dense or porous silica shells, both with and without electron–phonon coupling.}
    \label{fig:FixF_silicawatertemp}
\end{figure*}
This effect is particularly evident for the porous silica shells, where the reduced interfacial thermal conduction between gold and porous silica limits the rate of energy transfer. Our results reveal that, for 100 fs and 10 ps laser pulse durations, there exists a threshold time beyond which the silica shell dissipates heat to the surrounding water more efficiently. This enhanced dissipation occurs when the temperature of the dense silica shell approaches that of the gold core, thereby reducing the temperature difference that drives interfacial heat transfer. Therefore, the superior thermal conductance at the silica-water interface which is approximately 5 to 6 times higher than that at the water-gold interface dominates, leading to an accelerated cooling of the nanoparticle. While both the “+$\sigma_{es}$” and “–$\sigma_{es}$” cases exhibit this behavior, the time and magnitude of the temperature drop differ, highlighting the critical role of the interfacial electron-phonon coupling in determining the overall cooling rate.

Notably, this trend is not observed at low energy fluences (below $1$ mJ/cm\(^2\)) for a $1$ ns laser pulse. However, at higher fluences, the thermal response begins to follow the same trend observed with shorter laser pulses of 100 fs and 10 ps as shown in the Supplemental Material. The reduced energy deposition at low fluences leads to less pronounced thermal dynamics and a more subdued cooling response. This effect is further enhanced by the lower absorption efficiency of the dense silica shell nanoparticle compared to bare gold. 

Generally, our analysis indicates that the faster heat conduction observed for nanoparticles with thin silica shells can be attributed to the enhanced electron phonon coupling between the gold core and the silica shell \cite{xie-goldsilica}. This coupling facilitates efficient energy transfer, leading to rapid temperature equilibration and faster cooling. However, one must note that the experimental conditions in \cite{xie-goldsilica} employed very high laser fluences, which could result in extreme rises in water temperature at the nanoparticle interface potentially exceeding the solvent boiling point~\cite{gutierrez-varela2023} and may even cause fragmentation of the gold core. These considerations suggest that further investigation into optimal laser fluence levels is warranted to ensure nanoparticle stability and thermal performance.

Now, we calculate the electronic cooling rate of the three different nanoparticles. To quantify the rate at which the excited electron subsystem cools down, one must identify the maximum electron temperature, $T_{\mathrm{e,max}}$, and the time at which it occurs, $t_{\mathrm{max}}$. Next, the temperature is tracked until it decreases to half its maximum value, $T_{\mathrm{e,max}}/2$, at time $t_{1/2}$. The difference $\Delta t = t_{1/2} - t_{\mathrm{max}}$ thus represents the characteristic cooling time. The cooling rate, $R_{\mathrm{cool}}$, is then given by
$R_{\mathrm{cool}} = \frac{T_{\mathrm{e,max}}/2}{\,t_{1/2} - t_{\mathrm{max}}\,}$. 

We show the maximum electron temperature, $T_{\mathrm{e,max}}$ and half its maximum value, $T_{\mathrm{e,max}}/2$ in the electronic temperature plots of Fig.~\ref{fig:FixF_fspsns} and the cooling rate histograms in Fig.~\ref{fig:histocooling}.  We do not report the results without electron-phonon interaction between gold and silica, as they are nearly indistinguishable from those of bare gold as it can be seen from the electronic temperature plots of Fig.~\ref{fig:FixF_fspsns}. In the 100\,fs and 10\,ps cases, the electron temperature rises sharply following laser excitation, then decays rapidly. For the 100\,fs pulse, the dense silica nanoparticle exhibits the highest cooling rate (approximately 770\,K/ps), whereas the bare gold NP cools down at about 330\,K/ps, and the porous silica NP at around 630\,K/ps. Similar trends appear in the 10\,ps case, although the cooling rates (on the order of 250–350\,K/ps) are reduced due to more gradual energy deposition. By contrast, the 1\,ns pulse leads to markedly lower cooling rates (on the order of 0.2–0.3\,K/ps) for all nanoparticle types. Moreover, for the porous silica shell, the cooling is faster since the much slower energy deposition in gold under a ns pulse effectively minimizes the impact of electron–phonon coupling between gold and silica, giving rise to the enhanced interfacial thermal conductance between the porous silica and water.

Overall, dense silica nanoparticles exhibit the fastest cooling, due to their higher thermal conductivity and enhanced interfacial thermal conductance at the gold silica interface. Porous silica shells, by contrast, may confine more energy because of reduced conduction pathways, resulting in a slightly slower cooling rate. However, under a 1 ns laser pulse, porous silica shells show a relative advantage, as the enhanced interfacial heat transfer to water becomes more effective with the prolonged energy deposition. Nonetheless, both dense and porous shells exhibit faster cooling rates as compared to bare gold, underscoring the significance of the silica shell in governing electron cooling dynamics.

Fig.~\ref{fig:FixF_silicawatertemp} presents the simulation results of the temperature distribution at the nanoparticle water interface, computed at a fixed fluence of 0.1\,mJ/cm$^2$ and evaluated when the water temperature at the interface reaches its maximum value. The analysis is performed for the three laser pulse durations (100\,fs, 10\,ps, and 1\,ns), allowing a direct comparison of the maximum thermal response of the nanoparticles across different excitation regimes. Each panel of the figure is divided into four distinct regions corresponding to: the gold core temperature, the silica shell temperature (with a fixed thickness of 5\,nm), the temperature of water in direct contact with the nanoparticle (water interface), and the temperature of water located 2\,nm away from the interface. 
For all pulse durations, the highest temperature is observed in the gold core, with temperatures progressively decreasing toward the surrounding water. For the ultrashort 100\,fs pulse, the steep temperature gradient at the gold water interface is particularly pronounced, reflecting rapid energy deposition before significant thermal equilibration can take place. In contrast, for the longer 10\,ps and 1\,ns pulses, the extended timescales allow for more effective heat diffusion, resulting in a more uniform temperature distribution across the measured regions. Importantly, the temperature jump between the silica shell surface and water is consistently smaller than that observed between the gold core water. This reduced temperature difference highlights the higher interfacial thermal conductance between silica and water, which facilitates heat transfer into the surrounding medium. Additionally, porous shells, in particular, promote faster heat dissipation into water compared to dense shells, thereby lowering the peak temperature at the shell surface. In all cases, the water region at 2\,nm from the interface remains significantly cooler than the interface itself.

\section{Conclusion} \label{sec:conclusion}
In this work, we present a detailed investigation of the thermal response of gold core silica shell nanoparticles immersed in water, focusing on both dense and porous silica shells under laser fluences ranging from 0.05 to 5\,mJ/cm$^2$ and pulse durations of 100\,fs, 10\,ps, and 1\,ns. Our methodology accurately captures the key heat transfer mechanisms at the nanoscale by combining TTM with MD simulations and electronic temperature corrections to the Mie theory for gold. 

Our results demonstrate that, although increasing silica shell thickness generally increase the time required to heat the surrounding water, a thin dense silica shell (5\,nm) significantly enhances heat transfer compared to a bare gold nanoparticle for ultra-short laser pulses (fs - ps range laser pulse). This improvement arises from strong electron-phonon coupling at the gold-silica interface, which drives rapid temperature equilibration and facilitates efficient energy dissipation into the water. In contrast, thicker shells and lower interfacial coupling lead to a more gradual thermal response, emphasizing the importance of shell microstructure and thickness in controlling the nanoparticle heating behavior. 

Under ns laser pulse, substantial heat diffusion is observed. These results further indicate that for ns irradiation with low fluences (lower than 1 \,mJ/cm$^2$) bare gold nanoparticles act as more efficient heat generators than their silica coated counterparts; however, at higher fluences, the behavior follows the trend observed with ultra-short laser pulses. 

These findings offer critical insights into optimizing silica coated nanoparticles for photothermal and related applications~\cite{larquey2025}.
This study also calls for a systematic \textit{ab initio} investigation of the electron–phonon coupling at the gold silica interface to further elucidate the fundamental mechanisms governing energy transfer in illuminated core-shell nanoparticles.

\section*{Acknowledgement}
We thank interesting discussions with A. Crut and D. Amans. This work is supported by the ANR, France project "CASTEX" ANR-21-CE30-0027-01. 

\section*{Data availability}

The data that support the findings of this study are available from the corresponding author upon reasonable request. \\
The Python code used to compute the absorption efficiency using Mie theory, including temperature-dependent dielectric functions for gold, is available at:\\
\url{https://github.com/julienhajj/mie_absorption_gold/tree/main/mie_absorption_gold}
Access can be granted upon reasonable request to the corresponding authors.

\bibliography{bib}
\end{document}